\documentclass[review]{elsarticle}

\usepackage{hyperref}

\usepackage{xcolor}

\usepackage{amsmath}
\usepackage{amssymb}
\usepackage{amsfonts}
\usepackage{bm}
\usepackage{float}
\usepackage{import}
\usepackage{graphicx}

\usepackage{subcaption}

\usepackage[a4paper, left=3.0cm, right=3.0cm, top=3.0cm, bottom=3.0cm]{geometry}

\journal{Journal of Sound and Vibration}

\usepackage{numcompress}

\begin{document}

\begin{frontmatter}

\title{A low frequency model for the aeroacoustic scattering of cylindrical tube rows in cross-flow}

\author[imperial]{Aswathy~Surendran \fnref{fn1}}
\author[kth,marcus,linne]{Wei~Na} 
\author[imperial]{Charles~Boakes} 
\author[imperial]{Dong~Yang \fnref{fn2}}
\author[imperial]{Aimee~Morgans\corref{cor1}} 
\ead{a.morgans@imperial.ac.uk}
\address[imperial]{Department of Mechanical Engineering, Imperial College London, London, United Kingdom, SW72AZ}
\cortext[cor1]{Corresponding author}
\fntext[fn1]{Professur f\"{u}r Thermofluiddynamik, Technische Universität München, 85747 Garching, Germany}
\fntext[fn2]{Department of Mechanics and Aerospace Engineering, Southern University of Science and Technology, Shenzhen, 518055, PR China}

\author[kth,marcus,linne]{Susann~Boij} 
\address[kth]{Department of Engineering Mechanics, KTH Royal Institute of Technology, SE-10044, Stockholm, Sweden}
\address[marcus]{Marcus Wallenberg Laboratory for Sound and Vibration Research, Stockholm, Sweden}
\address[linne]{Linn\'{e} FLOW Centre, Stockholm, Sweden}

\begin{abstract}
Heat exchanger tube rows can influence the thermoacoustic instability behaviour of combustion systems since they act as both acoustic scatterers and unsteady heat sinks. Therefore, with careful tuning of their thermoacoustic properties, heat exchangers have the potential to act as passive control devices. In this work, we focus on (only) the acoustic scattering behaviour of heat exchanger tubes. We present a comparison of existing acoustic models for tube rows and slits, models for the latter having the advantage of incorporating frequency dependence. We then propose a new model that enables the adaptation of slit models for tube rows. This model is validated against experiments and Linearised Navier Stokes Equations (LNSE) predictions for the transmission and reflection coefficients, including phase information. The model predictions show very good agreement with the experimental and numerical validations, especially for low frequencies (Strouhal number $< 0.5$, based on tube radius and excitation frequency), with mean differences less than 2\% for the transmission coefficients (the reflection coefficient errors are somewhat larger since their magnitudes are very close to zero).
\end{abstract}


\end{frontmatter}

\section{Introduction\label{sec:intro}}
Heat exchangers are integral components of combustion systems like domestic boilers and industrial furnaces. Previous studies in domestic boilers \cite{Surendran2018_c,Surendran2018_d,Hosseini2017} have shown that heat exchangers, much like flames and other heat sources, can influence the thermoacoustic instability behaviour in these boilers. Thermoacoustic instability arises due to the interaction between acoustic fluctuations and unsteady heat release/absorption rates. Typically, the two-way interaction between the flame and the acoustic waves is considered. When this interaction develops into a positive feedback loop, the system experiences thermoacoustic instability which is characterised by large amplitude, low frequency self-excited pressure fluctuations that can be catastrophic \cite{Lieuwen2005_b,Poinsot2017}. In a system containing heat exchangers as well as or instead of flames, the thermoacoustic contribution of the heat exchangers must also be accounted for. This is assumed to be dominated by two effects: unsteady heat transfer across the heat exchanger and acoustic scattering at the heat exchanger. 

\begin{figure}[H]
	\centering
	\includegraphics[width=0.65\columnwidth]{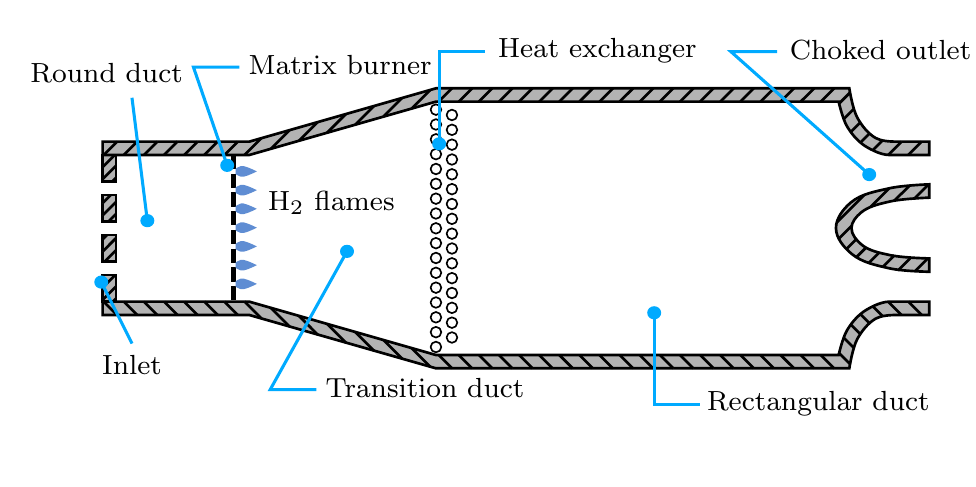}
	\caption{Schematic of a preburner combustor with heat exchanger.\label{fig:sabre}}
\end{figure}

Assuming that the heat exchanger comprises of rows of cylindrical tubes, the first effect - the unsteady sink (or source depending on the operating conditions \cite{Hosseini2017}) - responds to oncoming acoustic fluctuations. The unsteady heat transfer then acts as a monopole acoustic source, further contributing to acoustic fluctuations. The second effect - the acoustic scattering at the tube row - is primarily due to the geometry and flow, and can be studied initially for a cold flow with no heat transfer present. In the absence of a mean flow, the acoustic waves are scattered by the tube rows in which viscous effects and thereby attenuation or damping is normally small at the low frequencies relevant to thermoacoustics \cite{SurendranInt2014, Heckl2019}. On the other hand, in the presence of a mean flow, flow separation and vortex shedding can significantly enhance sound attenuation \cite{Howe1998_book}. However, a strict analytical solution for acoustic scattering in tube rows with mean flow is very challenging due to the complex flow structure, vortex shedding and coupling with acoustics \cite{Hong2020}. It is important to accurately model these two effects. The present work is dedicated to the acoustic scattering effect of heat exchanger tube rows (without heat transfer), and is motivated by the use of heat exchangers in futuristic aero-propulsion engines as shown in Figure~\ref{fig:sabre}. 

\vspace{-2mm}

Heat exchanger tubes are typically circular in cross-section. Surendran et al.~\cite{Surendran2018_b} developed the quasi-steady model for the acoustic scattering of tube rows in cross-flow. Though the model predictions agreed well with experimental results for low Strouhal and low Mach number flows, they do not account for any frequency dependence and therefore lack information about the phase of the acoustic response. Phase information is important for thermoacoustic systems as it determines the formation of the positive feedback loop between acoustic fluctuations and heat release/absorption rate fluctuations. At the low frequencies at which thermoacoustic instabilities occur, the diameter of the tubes is greatly exceeded by the acoustic wavelength, and when the tubes are in a ``cross-flow'' arrangement, it is possible to approximate them as thin plates of rectangular cross-section, separated by a rectangular gap. Such geometries appear as ``slits'' to the incoming flow. While there are very few aeroacoustic models for a row of cylindrical tubes, the aeroacoustic scattering effect of slits and perforated plates has been more widely studied \cite{Lahiri2017}. Studies conducted with slit configurations have been shown to have a stabilising influence on combustion systems undergoing thermoacoustic instabilities \cite{Surendran2017_b, Albaharna2018}. Instead of slit plates, Quinn and Howe \cite{Quinn1984} approximated a row of rigid circular tubes with a row of infinitesimally thin rigid horizontal strips. They found that acoustic attenuation at the row increases with decreasing Strouhal number i.e., for a given tube diameter and frequency, attenuation increases with increasing cross-flow velocity.   

In the present work, we focus on two slit models: Dowling and Hughes slit model \cite{Dowling1992} applied to rectangular slits and modified Cummings slit model \cite{Albaharna2018}. The predictions for the acoustic scattering matrices from these models are compared in order to devise a method which will aide in adapting slit models for describing the scattering behaviour of tube rows. The advantage of using slit models is that they account for frequency dependence (at low frequencies) which is missing in existing literature for cylindrical tube rows in cross-flow.

The structure of the paper is as follows: three existing acoustic scattering models which are relevant to the tube array scattering problem are described in Section~\ref{sec:models}. These are the quasi-steady cylinder model for an array of cylindrical tubes \cite{Surendran2018_b} as well as the Dowling and Hughes \cite{Dowling1992} and modified Cummings \cite{Albaharna2018} models, both for an array of rectangular slits. In Section~\ref{sec:mod_compare}, we compare the acoustic scattering matrix predictions from the three models, as applied to a row of cylindrical tubes  and slits of similar spacing. This allows us to draw preliminary conclusions for tube rows, enabling us to improve and adapt the modified Cummings slit model for tube row scattering predictions at low Strouhal numbers, as described in Section~\ref{sec:kadj_mc_model}. The proposed model is validated against experimental measurements and numerical simulations for arrays of cylindrical tubes described in Section~\ref{sec:validation}. The validation results are shown in Section~\ref{sec:results} and conclusions drawn in Section~\ref{sec:conclusion}. 

\section{Acoustic models considered\label{sec:models}}
The cross-flow past circular heat exchanger tubes has similar flow dynamics to that of the flow through a gap, a constriction or a slit i.e., they all produce jets and have associated shear layers downstream of the gaps. When a low frequency acoustic wave is incident on any of these geometries, it creates in the nearfield higher order evanescent modes and planar propagating waves. The evanescent modes decay faster than the planar waves, and are typically found only close to the gap. As we move farther from the constriction, only low frequency planar waves, that are one dimensional, persist. In most combustion systems undergoing thermoacoustic instabilities, the frequency range of interest is generally $\sim \mathcal{O}(100)$Hz. In the present study, we aim to develop an aeroacoustic model for tube rows subjected to low Mach number flows and low frequency excitations. To this end, we compare some of the existing aeroacoustic models as applied to tube rows, identify their limitations and then use this information to build our new model. We consider the following models that account differently for the gap shapes.
\begin{itemize}
	\item A quasi-steady model developed for a row of cylinders (Section~\ref{sec:qsmodel})
	\item The Dowling and Hughes model for thin slits (Section~\ref{sec:dhmodel})
	\item The modified Cummings model developed for thin slits (Section~\ref{sec:mcmodel})
\end{itemize}

\subsection{Quasi-steady cylinder model\label{sec:qsmodel}}
\begin{figure}[H]
	\centering
	\includegraphics[width=0.55\columnwidth]{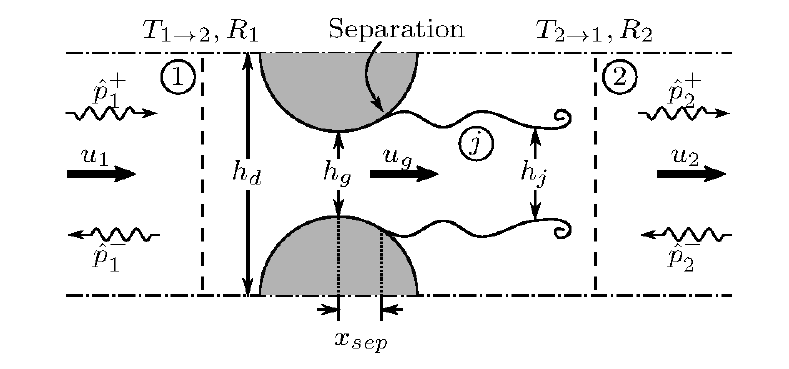}
	\caption{Schematic for the flow domain in the Quasi-steady cylinder model\label{fig:QS}}
\end{figure}

The tube row geometry considered is illustrated as a schematic in Figure~\ref{fig:QS}. The geometry consists of two half cylinders placed inside a duct of height $h_d$ with the gap between them being $h_g$. A mean flow, known as bias flow, passes through the gap with velocity $u_g$. The bias flow then forms a jet of height $h_j$ downstream of the cylinder. It is sufficient to model this system instead of an array of cylinders as they are equivalent (through the method of image sources).

The quasi-steady model for tube rows was developed and experimentally validated by Surendran et al.~\cite{Surendran2018_b}. They assume that the time-dependent acoustic perturbations of the flow are sufficiently slow in order for the unsteady terms in the relevant 1-D conservation equations to be neglected, as first proposed by Ronneberger~\cite{Ronneberger1967}. Such an assumption is valid for small Strouhal numbers $(St_{QS}=f r /u_g)$  and small Helmholtz numbers $(He=2\pi f d/c)$, where $f$ is the frequency in Hz, $d$ is the diameter of the tube, $r$ is the radius of the tube, $u_g$ is the gap velocity and $c$ is the speed of sound. Small $He$ requires the acoustic wavelength to be much larger than the tube diameter, leading to a {\em compact} tube row assumption. Furthermore, it is assumed that the acoustic wavelength greatly exceeds the spatial extent of the mixing zone downstream of the tube row. Consequently, the acoustic scattering properties are expected to have {\em no phase changes} across the tube row.

In the quasi-steady model, the duct containing the half cylinders is divided into three regions: Region 1, region $j$ and region 2 (see Figure ~\ref{fig:QS}). Regions 1 and 2 have uniform flows and are upstream and downstream of the cylinders, respectively. Region $j$ lies between regions 1 and 2 and contains the cylinders, the jet and the mixing region. Assuming isentropic and irrotational flow between the regions 1 and $j$, the conservation of mass and energy under the assumptions of quasi-steadiness and perfect gas behaviour can be applied. This yields Eqs.~\eqref{eq:qsR1jmass}-\eqref{eq:qsR1jisen}, where the variables $\rho$, $p$ and $\gamma$ denote the density, pressure and ratio of specific heats, respectively, while the subscripts indicate the corresponding regions. 
\begin{align}
\label{eq:qsR1jmass}
&&h_d\rho_1 u_1 &= h_j \rho_j u_j \: &&\left(\text{continuity}\right)&&\\
\label{eq:qsR1jenergy}
&&\frac{1}{2}u_1^2 + \frac{\gamma}{\gamma -1}\frac{p_1}{\rho_1} &= \frac{1}{2}u_j^2 + \frac{\gamma}{\gamma -1}\frac{p_j}{\rho_j} \: &&\left(\text{energy}\right)&&\\
\label{eq:qsR1jisen}
&&\frac{p_1}{p_j} &= \left(\frac{\rho_1}{\rho_j}\right)^\gamma \: &&\left(\text{isentropic}\right)&&
\end{align}
Downstream of the mixing region, the flow is assumed to be uniform and non-isentropic. Here, the conservation of mass and momentum are used to link regions $j$ and 2, as shown in Eqs.~\eqref{eq:qsRj2mass}-\eqref{eq:qsRj2mom}. Neglecting heat transfer, viscous and frictional losses at the wall, conservation of energy can be applied across regions 1 and 2 as shown in Eq.~\eqref{eq:qsR12energy}.
\begin{align}
\label{eq:qsRj2mass}
&&h_j \rho_j u_j &= h_d \rho_2 u_2 \: &&\left(\text{continuity}\right)&&\\
\label{eq:qsRj2mom}
&&h_dp_j + h_j \rho_j u_j^2 &= h_dp_2 + h_d\rho_2 u_2^2\: &&\left(\text{momentum}\right)&&\\
\label{eq:qsR12energy}
&&\frac{1}{2}u_1^2 + \frac{\gamma}{\gamma -1}\frac{p_1}{\rho_1} &= \frac{1}{2}u_2^2 + \frac{\gamma}{\gamma -1}\frac{p_2}{\rho_2} \: &&\left(\text{energy}\right)&&
\end{align}

Equations ~\eqref{eq:qsR1jmass}-\eqref{eq:qsR12energy} are then linearised about a mean condition by decomposing the variable $u$, $p$ and $\rho$ into sums of mean $\left(\bar{~}\right)$ and small acoustic perturbation components $\left('\right)$, while neglecting the higher order terms of the primed quantities \cite{Dowling2003}. These decompositions are shown in Eq.~\eqref{eq:decomposed} and the derivation of the linearised equations can be found in Appendix D of \cite{Surendran_thesis}. The acoustic perturbations can be represented in terms of the complex amplitudes of the forward and backward travelling pressure waves $\hat{p}_{1,2}^+$ and $\hat{p}_{1,2}^-$ as shown in Eq.~\eqref{eq:perturbation}, where $s'$ is the entropy perturbation, $c_p$ is the specific heat at constant pressure and the subscript $i$ denote the region. Substituting these expressions into the linearised conservation equations yields a set of equations in terms of $\hat{p}_{1,2}^\pm$ that can be manipulated to form the scattering matrix. In the absence of an incoming entropy wave ($s'_1=0$), the scattering matrix will be of the form shown in Eq.~\eqref{eq:qs_scat}. Here, $T_{1\rightarrow2}$ and $R_1$ are the transmission and reflection coefficients for a wave incident from the upstream section and $T_{2\rightarrow1}$ and $R_2$ are the transmission and reflection coefficients for a wave incident from the downstream section.
\begin{align}
\label{eq:decomposed}
&&p &= \bar{p}+p' \; &&u = \bar{u} + u' \; &&\rho = \bar{\rho} + \rho'&&\\
\label{eq:perturbation}
&& p_i ' &= \hat{p}_i^+ + \hat{p}_i^- \; &&u_i ' = \frac{\hat{p}_i^+ - \hat{p}_i^-}{\bar{\rho}_i c_i} \; &&\rho_i ' =\frac{\hat{p}_i^+ + \hat{p}_i^-}{c_i^2}-\frac{\bar{\rho}_i}{c_p}s_i'&&
\end{align}
\begin{equation}
\label{eq:qs_scat}
\left[\begin{matrix}
\hat{p}_2^+ \\ \hat{p}_1^-
\end{matrix}\right] = \left[\begin{matrix}
T_{1\rightarrow 2} & R_2 \\ R_1 & T_{2\rightarrow1}
\end{matrix}\right]\left[\begin{matrix}
\hat{p}_1^+ \\ \hat{p}_2^-
\end{matrix}\right]
\end{equation}

\subsection{Dowling and Hughes slit model\label{sec:dhmodel}}
\begin{figure}[H]
	\centering
	\includegraphics[width=0.55\columnwidth]{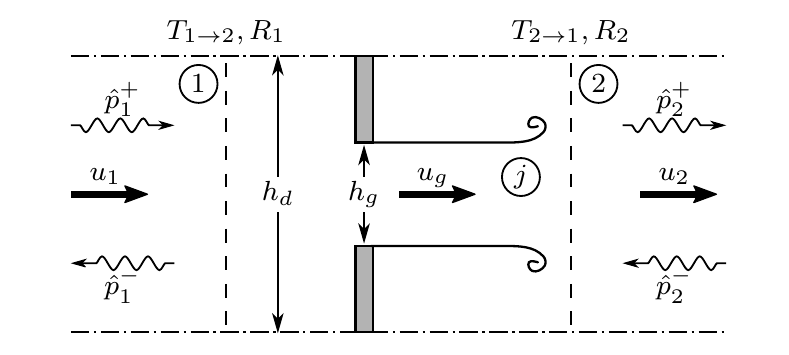}
	\caption{Schematic for the flow domain in the Dowling and Hughes slit model.}
	\label{fig:DH}
\end{figure}

The Dowling and Hughes model \cite{Dowling1992} describes the acoustic scattering behaviour of thin rectangular slits with bias flow, by extending the analysis of Howe~\cite{Howe1979b}. The flow domain consists of two infinite half planes connected by  an infinitely long slit plate. The slit plate, as shown in Figure~\ref{fig:DH}, can be approximated to the geometry shown, again through the method of image sources. Each slit has a width of $h_g$ and is spaced $h_d$ distance apart, having a bias flow velocity of $u_g$ though the slits. The flow is assumed to be incompressible and hence $\bar{\rho}_1 = \bar{\rho}_2 = \rho_0$.

In the limit of small open area ratio $\left(\eta=h_g/h_d\right)$ and small Helmholtz number, and solving the inhomogeneous Helmholtz equation in stagnation enthalpy, the transmission and reflection coefficients can be determined from Eqs.~\eqref{eq:dhT} and \eqref{eq:dhR}.
\begin{align}
\label{eq:dhT}
T_{1\rightarrow2}&=\rho_0\omega \dot{V}/\left(k h_d\right),\\
\label{eq:dhR}
R_{1} &= 1 - T_{1\rightarrow2},
\end{align}
with
\begin{equation}
\label{eq:dh_vol}
\frac{\rho_0\omega\dot{V}}{k h_d} = \frac{i\pi\eta/\left(2StM_g\right)}{i\pi\eta/\left(2StM_g\right)-\ln{\left(\pi\eta\right)} +\ln{2}/\Phi(St)},
\end{equation}
\begin{equation}
\label{eq:dh_phi}
\Phi(St) = 1 - \frac{1}{St \ln 2}\left\lbrace \frac{\pi I_0\left(St\right)e^{-St} + 2i \sinh\left(St\right)K_0\left(St\right)}{\pi e^{-St}\left[I_1\left(St\right)+\frac{I_0\left(St\right)}{St \ln 2}\right] + 2i\sinh\left(St\right)\left[\frac{K_0\left(St\right)}{St \ln 2}\right]-K_1\left(St\right)} \right\rbrace,
\end{equation}
where $\dot{V}$ is the perturbation volume flux through the slit, $St = \omega h_g/(2u_g)$, $M_g=\bar{u}_g/c$ is the mean Mach number of the bias flow and $I_m$ and $K_m$ are the modified Bessel functions of order $m$. Again, the scattering matrix would be of the form shown in Eq.~\eqref{eq:qs_scat}.

\subsection{Modified Cummings model for slits \label{sec:mcmodel}}
\begin{figure}[htbp!]
	\centering
	\begin{subfigure}[b]{0.49\textwidth}
		\centering
		\includegraphics[width=\columnwidth]{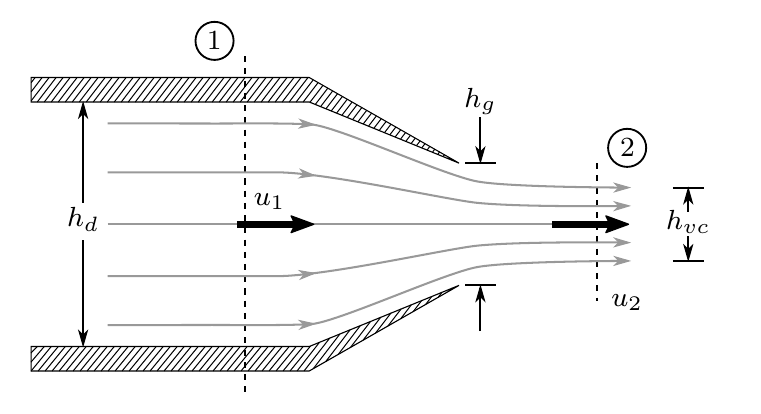}
		\caption{Converging nozzle flow (Cummings and Eversman~\cite{Cummings1983})}
		\label{fig:cummings}
	\end{subfigure}
	\hfill
	\begin{subfigure}[b]{0.49\textwidth}
		\centering		
		\includegraphics[width=\columnwidth]{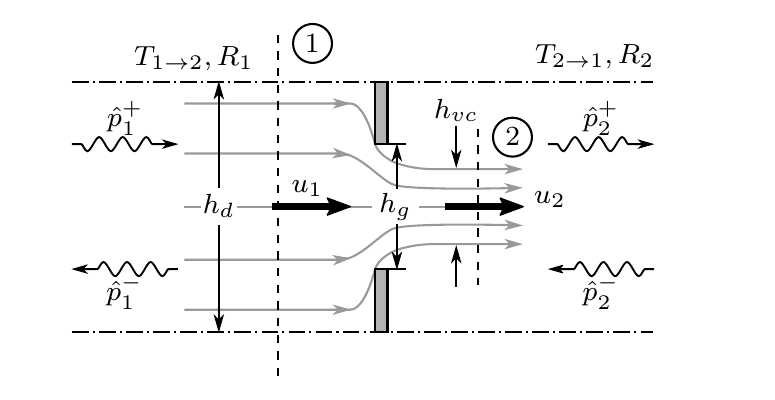}
		\caption{Flow though a slit (Present study)}
		\label{fig:MC}
	\end{subfigure}
	\caption{Geometry used in the modified Cummings model. \label{fig:mod_cummings}}
\end{figure}

Cummings~\cite{Cummings1983,Cummings1986} developed a model to describe the acoustic transmissions through duct terminations using the unsteady Bernoulli equation. The configuration studied in \cite{Cummings1983} is shown in Figure~\ref{fig:cummings}; an incompressible uniform mean flow with superimposed plane acoustic waves travels through a converging nozzle of circular cross-section. The flow then separates at the nozzle exit to form a jet with diameter $h_{vc}$ ({\em vena contracta}). In the present study, we adapt a similar flow structure for the slit (see Figure~\ref{fig:MC}), where the uniform flow upstream of the slit separates at the slit edge forming a jet. The velocities at 1 and 2 can be related to the gap velocity $u_g$ through the open area ratio $\eta$ ($= h_g/h_d$) and the contraction coefficient $\sigma = h_{vc}/h_g$ as $u_1 = \eta u_g$ and $u_2 = u_g/\sigma$.

\noindent Applying unsteady Bernoulli's equation from region 1 to 2 yields
\begin{equation}
\label{eq:cm_bernoulli}
\int_{1}^{2}\rho_0\frac{du_g}{dt} dx + \frac{\rho_0}{2}\left(u_2^2 - u_1^2\right)=p_1 - p_2. 
\end{equation}
Defining an effective length $L$ for the slug of fluid undergoing unsteady motion at the slit as $\int_{1}^{2}\rho_0 \left(du_g/dt\right)dx = \rho_0 L \left(du_g/dt\right)$, the Cummings equation (Eq.~\eqref{eq:cm_bernoulli}) can be written as
\begin{equation}
\label{eq:cm_eqn}
\rho_0 L \frac{du_g}{dt} + \rho_0 u_g^2 \frac{1-\eta^2\sigma^2}{2\sigma^2}=p_1 - p_2. 
\end{equation}

Cummings~\cite{Cummings1986} argued that for high amplitude pressure fluctuations, the jet formed at the orifice can grow and diminish periodically due to flow reversal, causing $L$ to vary. However, in our analysis, we restrict ourselves to low amplitude pressure fluctuations so that we do not encounter flow reversal at the slit. After representing Eq.~\eqref{eq:cm_eqn} in terms of the mean or steady and perturbed quantities and subtracting the steady contribution and linearising, one can obtain the equation for perturbations for small $\eta$ as \cite{Luong2005} 
\begin{equation}
\label{eq:mceqn_luong}
L\frac{du_g'}{dt} + \frac{u_g'}{\sigma^2}\left(\bar{u}_g+\frac{u_g'}{2}\right)=\frac{p_I}{\rho_0},
\end{equation}
where $p_I = p_1'-p_2'$. Albaharna~\cite{Albaharna2018} linearised Eq.~\eqref{eq:mceqn_luong} by neglecting the higher order terms i.e., products of perturbation terms, and evaluated the transmission and reflection coefficients. This linearised equation (Eq.~\eqref{eq:mceqn}) is referred to as the {\em modified Cummings model} in the present paper. 
\begin{equation}
\label{eq:mceqn}
L\frac{du_g'}{dt} + \frac{u_g'\bar{u}_g}{\sigma^2}=\frac{p_I}{\rho_0}
\end{equation}
Manipulating the results in \cite{Albaharna2018} for an acoustic wave incident from the upstream side with temporal variation $e^{i\omega t}$ (see Figure~\ref{fig:MC}), we obtain the transmission and reflection coefficients as
\begin{align}
\label{eq:mc_T12}
T_{1\rightarrow2} &= \frac{2}{1+\eta M_g}\left(\frac{4i\chi St M_g^2 +\left(M_g/\sigma\right)^2 +1}{4i\chi St M_g/\eta + M_g/\left(\eta \sigma^2\right)+2/\left(1+\eta M_g\right)}\right),\\
\label{eq:mc_R1}
R_{1} &= \frac{4i\chi St M_g/\eta +M_g/\left(\eta\sigma^2\right)}{4i\chi St M_g/\eta + M_g/\left(\eta \sigma^2\right)+2/\left(1+\eta M_g\right)},\\
T_{2\rightarrow 1} &= \frac{2/\left(1+\eta M_g\right)}{2/\left(1+\eta M_g\right) + M_g/\left(\eta \sigma^2\right) + 4i\chi St M_g/\eta},\\
R_{2} &= \frac{M_g/\left(\eta \sigma^2\right) + 4i\chi St M_g/\eta}{1 + \left(\frac{1+\eta M_g}{1 - \eta M_g}\right)\left(1 + M_g/\left(\eta \sigma^2\right) + 4i\chi St M_g/\eta\right)}
\end{align}
where $St = \omega h_g/(2u_g)$ and $\chi = L/(2h_g)$ is the end correction coefficient.

\subsubsection{Evaluation of $L$}
For low amplitude pressure perturbations that do not lead to flow reversal, Luong et al.~\cite{Luong2005} suggested that $L=2l_0+l_w$, where $l_0$ is the end correction on one side and $l_w$ is the thickness of the orifice. For thin orifices and slits, $l_w$ can be neglected.\\
Albaharna \cite{Albaharna2018} used the end correction model for rectangular perforations suggested by Vigran \cite{Vigran2014}, to define $\chi$. However, this model does not hold for situations where the slit length greatly exceeds slit width as $\chi$ becomes infinite. Therefore, in the present study, we use the end correction for a rectangular orifice in a baffle wall as given in Page 319 of \cite{Mechel2008_book} i.e.,
\begin{equation}
\label{eq:endcorrec}
\frac{l_0}{h_g} = \frac{1}{\pi}\ln\left[\frac{1}{2}\tan\left(\frac{\pi\eta}{4}\right)+\frac{1}{2}\cot\left(\frac{\pi\eta}{4}\right)\right]
\end{equation}

\section{Comparison of Models\label{sec:mod_compare}}
The transmission, reflection and absorption coefficients for the different models (using different geometries) are compared in this section. The absorption coefficient ($\Delta$) is defined as the ratio of the acoustic energy absorbed to the acoustic energy incident \cite{Morfey1971}. Across any two regions denoted by 1 and 2, $\Delta_{1\rightarrow2}$ can be expressed as
\begin{equation}
\label{eq:abscoeff}
\Delta_{1\rightarrow2} = 1-\left[\frac{\left|\hat{p}_1^-\right|^2\left(1-M_1\right)^2 + \left|\hat{p}_2^+\right|^2\left(1+M_2\right)^2\left(\bar{\rho}_1c_1/\bar{\rho}_2c_2\right)\mathcal{A}_2/\mathcal{A}_1}{\left|\hat{p}_1^+\right|^2\left(1+M_1\right)^2 + \left|\hat{p}_2^-\right|^2\left(1-M_2\right)^2\left(\bar{\rho}_1c_1/\bar{\rho}_2c_2\right)\mathcal{A}_2/\mathcal{A}_1}\right],
\end{equation}
where $\mathcal{A}$ is the area of cross-section and the notation $\Delta_{1 \to 2}$ denotes the absorption coefficient for an acoustic wave passing from region 1 to 2. For the analysis provided in this paper, we have used the properties shown in Table~\ref{tab:sys_prop}. Here, $d$ is the diameter of the cylinder and $\mu$ is the dynamic viscosity.
\begin{table}[h]
	\caption{System properties}
	\label{tab:sys_prop}
	\begin{center}
		\begin{tabular}{ccc|ccc}
			\hline 
			Property & Unit & Value & Property & Unit & Value\\ \hline
			$d$ & [mm] & 11.8 & $\mu$ & [Pa s] & 1.8$\times10^{-5}$\\ 
			$\rho$ & [kg/m$^3$] & 1.2 & $\eta$ & - & 0.31\\ 
			$c$	& [m/s] & 340 & $\gamma$ & - & 1.4\\  \hline
		\end{tabular} 
	\end{center}
\end{table}

\begin{figure}[H]
	\centering	
	\includegraphics[width=0.7\columnwidth]{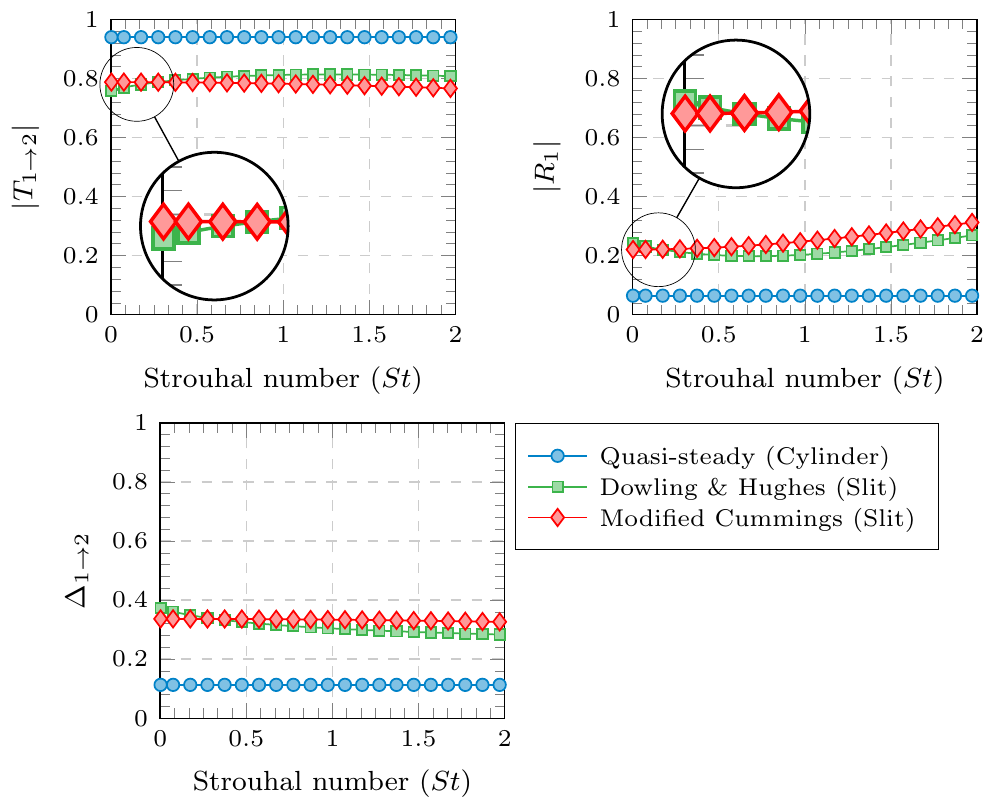}
	\caption{Transmission, reflection and absorption coefficients for an upstream perturbation at $M_g=0.1$, $\eta = 0.31$ and $\sigma=0.75$.}
	\label{fig:scat_all_st}
\end{figure}

Figure~\ref{fig:scat_all_st} shows the coefficients predicted by the three models for different excitation frequencies (in terms of the $St$), $M_g=0.1$, $\eta=0.31$ and $\sigma=0.75$. The Dowling and Hughes model (D\&H) and the modified Cummings model (MC) are for thin rectangular slits ($l_w=0$) and the Quasi-steady model (QS) is for tube rows. It can be observed that the D\&H and MC models are consistent and exhibit similar behaviour in frequency, whereas the QS model, experimentally validated in \cite{Surendran2018_b}, does not exhibit any dependence on frequency. It can also be noted that at very low Strouhal numbers $\left(St\leq0.5\right)$, the D\&H and MC models also predict quasi-steady behaviour i.e., the coefficients remain nearly constant, reinforcing the low Strouhal and low Helmholtz numbers assumptions used in the QS model for tube rows. However, the QS model predicts a higher transmission and lower reflection and absorption coefficients compared to the other two models.  

\begin{figure}[H]
	\centering
	\includegraphics[width=0.7\columnwidth]{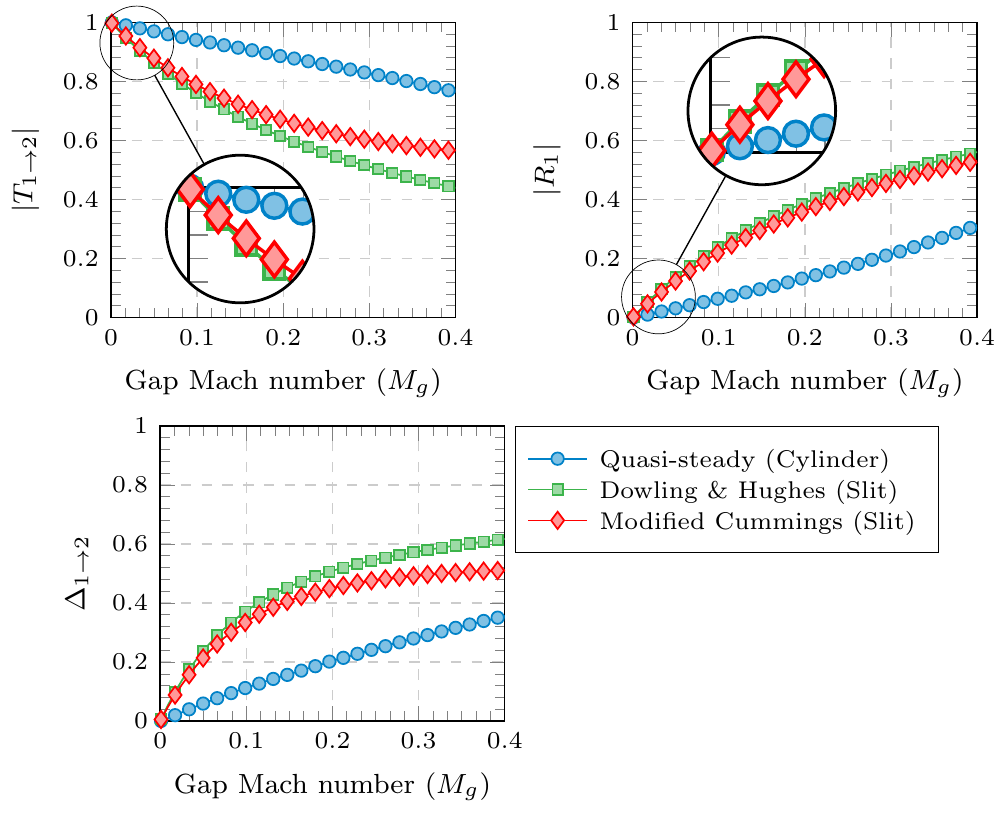}
	\caption{Transmission, reflection and absorption coefficients for an upstream perturbation at $St\rightarrow0$, $\eta = 0.31$ and $\sigma=0.75$.}
	\label{fig:scat_all_mach}
\end{figure}

For very low $St$ values, say $St\rightarrow 0$ ($= 10^{-6}$), Figure~\ref{fig:scat_all_mach} shows the variation of transmission, reflection and absorption coefficients with varying gap Mach number $M_g$ for $\eta = 0.31$ and $\sigma=0.75$. Though the overall trend exhibited by the different coefficients are similar, the QS model is seen to again predict higher transmission and lower reflection coefficients for lower $M_g$ values. As $M_g$ increases, the MC model is seen to diverge from the D\&H model for transmission and absorption coefficients, though the reflection coefficient is consistent. Since the QS model has been experimentally verified for tube rows~\cite{Surendran2018_b} and the D\&H model has been experimentally verified for plates with slits~\cite{Dowling2003}, it can be concluded that tube rows cannot be approximated as rectangular slits as was proposed in~\cite{Albaharna2018}. 

\section{Loss coefficient ($K$) adjusted Modified Cummings model\label{sec:kadj_mc_model}}
\begin{figure}[H]
	\centering
	\includegraphics[width=0.75\columnwidth]{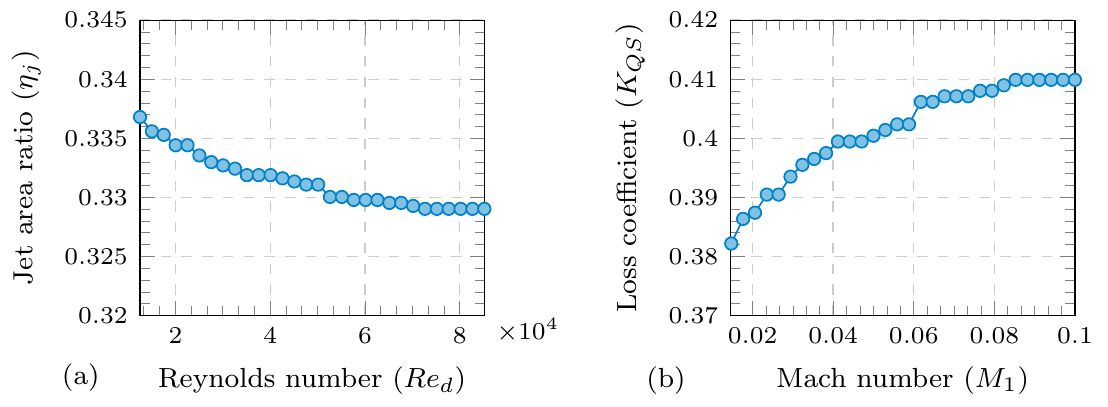}
	\caption{Variation of (a) jet area ratio ($\eta_j$) with Reynolds number ($Re_d$) and (b) loss coefficient ($K_{QS}$) with Mach number ($M_1$)}
	\label{fig:etaj_losscoeff_mach}
\end{figure}

\begin{figure}[htbp!]
	\centering
	\includegraphics[width=0.7\columnwidth]{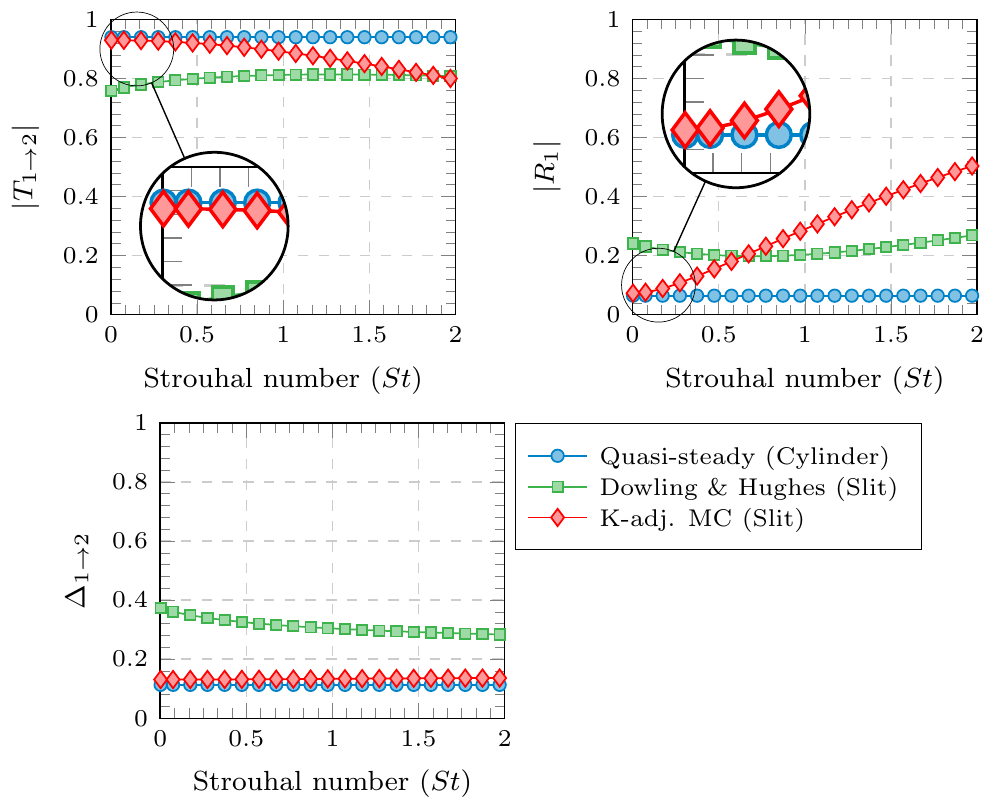}
	\caption{Transmission, reflection and absorption coefficients for an upstream perturbation for $M_g=0.1$ and $\eta = 0.31$. $\sigma$ in the MC model is adjusted to match $K_{MC}$ with $K_{QS}$.}
	\label{fig:scat_all_st_mod}
\end{figure}

\begin{figure}[htbp!]
	\centering
	\includegraphics[width=0.7\columnwidth]{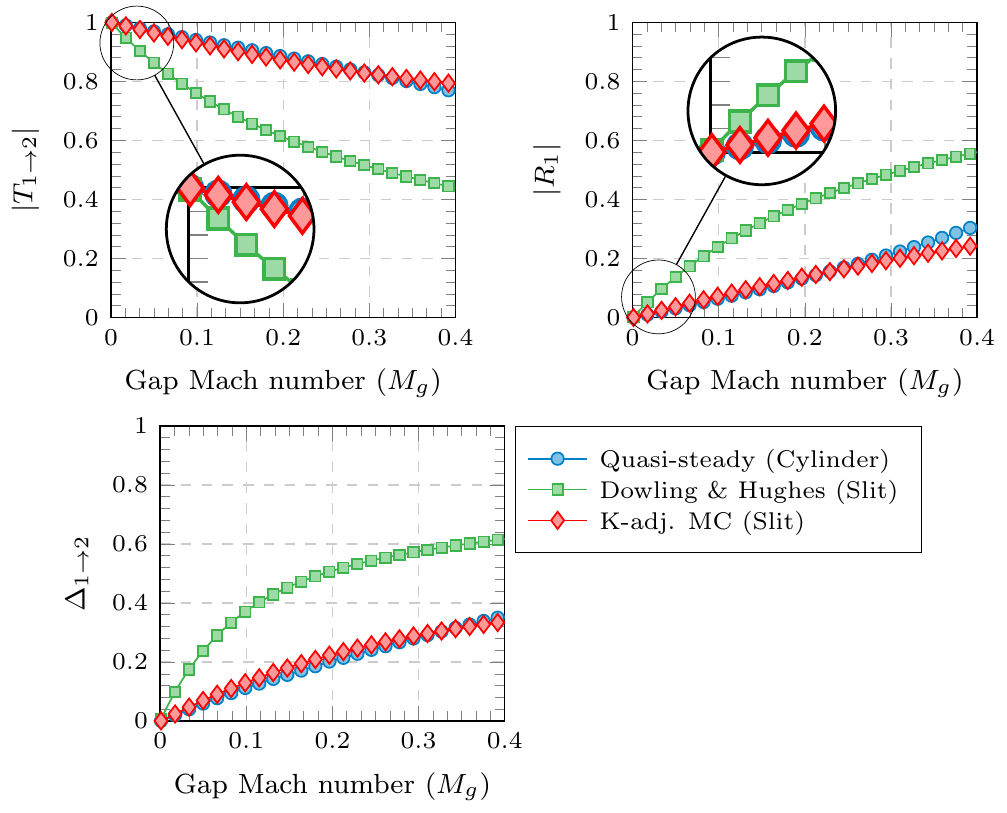}
	\caption{Transmission, reflection and absorption coefficients for an upstream perturbation for $St\rightarrow0$ and $\eta = 0.31$. $\sigma$ in the MC model is adjusted to match $K_{MC}$ with $K_{QS}$.}
	\label{fig:scat_all_mach_mod}
\end{figure}

The discrepancy between the predictions of the QS and MC models can be explained by the different incompressible steady flow loss coefficients implied by Eq.~\eqref{eq:qsRj2mom} and Eq.~\eqref{eq:cm_eqn}. The loss coefficient is defined as 
\begin{equation}
\label{eq:losscoeff}
K = \frac{\Delta p_0}{0.5\rho_0u_g^2},
\end{equation}
where $\Delta p_0$ is the stagnation pressure difference between locations 1 and 2. For the QS model (refer to Figure~\ref{fig:QS}) and the MC model (refer to Figure~\ref{fig:MC}), the loss coefficients across 1 and 2 are evaluated as\vspace{-4mm}
\begin{align}
\label{eq:lossqs}
K_{QS} &= \left(\eta/\eta_j\right)^2\left(1-\eta_j\right)^2,\\
\label{eq:lossmc}
K_{MC} &= \left(\frac{1}{\sigma^2}-\eta^2\right).
\end{align}
where $\eta_j = h_j/h_d$. The variation of $\eta_j$ with incoming Reynolds number $Re_d=\rho u_g d/\mu$ and $K_{QS}$ with incoming Mach number $M_1$ are shown in Figure~\ref{fig:etaj_losscoeff_mach}. As $Re_d$ increases, the separation location moves closer to the throat i.e., $\eta_j$ decreases and therefore $K_{QS}$ increases. Since $\eta$, $\eta_j$ and $\sigma$ are less than 1, $K_{MC} > K_{QS}$, demonstrating that the MC model has greater stagnation pressure loss or dissipation than the QS model. In order for the slit to represent the scattering behaviour of tube row, we treat $\sigma$, the contraction coefficient, as an arbitrary quantity and match $K_{MC}$ with $K_{QS}$. This results in $\sigma$ being greater than 1 and points to the jet expanding after separation at the gap. Additionally, we have also assumed that the slit plate is thick with thickness $l_w = 2 x_{sep}$, where $x_{sep}$ is the displacement between the throat and the separation location for the cylindrical tube row, as shown in Figure~\ref{fig:QS}. $x_{sep}$ can either be evaluated analytically or numerically from simulations; in the present analysis, we use Thwaites method \cite{Kundu2012_b} to compute it. The effect of matching the loss coefficients on the acoustic scattering behaviour is shown in Figures~\ref{fig:scat_all_st_mod} and~\ref{fig:scat_all_mach_mod}. The low Strouhal limit of the MC model now coincides with the QS model. The small deviation between the two models at higher Mach numbers in Figure~\ref{fig:scat_all_mach_mod} may be due to the different linearised forms of Eqs.~\eqref{eq:qsRj2mom} and \eqref{eq:cm_eqn}. These figures show that the loss coefficient adjusted MC model can be used to predict the scattering behaviour of tube rows or other complex geometries, if an appropriate contraction ratio ($\sigma$) and end corrections ($l_0$ and $l_w$) are applied.

\section{Validation\label{sec:validation}}
We have seen that the proposed loss coefficient ($K$) adjusted MC model recovers the quasi-steady cylinder row model in the zero frequency limit. A key benefit of the $K$-adjusted MC model is that it predicts frequency dependence. We now seek to validate the frequency dependence that it predicts using numerical simulations and experiments.

\subsection{Numerical Simulations\label{sec:num_simulation}}
\begin{figure}[H]
	\centering
	\includegraphics[width=0.75\columnwidth]{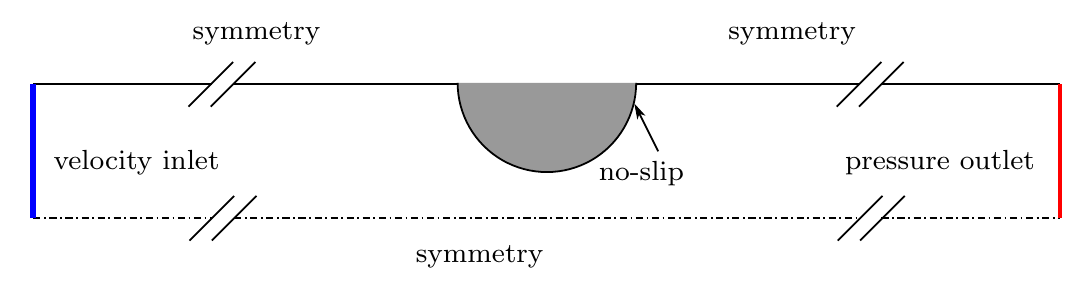}
	\caption{Boundary conditions used in numerical simulations.}
	\label{fig:comsol}
\end{figure}

Linearised Navier-Stokes Equations (LNSE) method described in \cite{Kierkegaard2010, Surendran2019_c} is used to numerically validate the proposed model. This is a two step process where the steady field is simulated first using a compressible Reynolds-Averaged Navier-Stokes (RANS) solver with SST turbulence model. Next the perturbed field including acoustic and vorticity perturbations, perturbed around the steady state, is computed from the linearised Navier-Stokes equations in the frequency domain. Once the two fields are computed, the scattering matrix of the sample is evaluated at each frequency using a two-source method \cite{Boden1986, Abom1991}. The computations for both steady flow field and perturbed flow field are carried out in COMSOL Multiphysics V5.3. The flow domain, whose schematic is shown in Figure~\ref{fig:comsol}, consisted of only half the geometry i.e., a domain with one half cylinder (top) and centreline (bottom) along with the boundary conditions of (i) uniform velocity condition at the inlet (left), (ii) symmetric condition on the top and bottom of the domain, (iii) no-slip condition at the cylinder wall and (iv) ambient pressure boundary condition at the outlet (right). The boundary conditions and the flow domain are the same for both steady and acoustic simulations except that in the acoustic mesh, there are additional buffer zones at the inlet and outlet known as PML zones (perfect match layer) that help to minimise acoustic reflections. After grid independence study, we have chosen a RANS simulation mesh with $\sim$140000 elements, both triangular and quadrilateral (close to the cylinder). The maximum element size is $2.9\times 10^{-4}$m and the minimum element size is $4.2\times 10^{-6}$m. In the flow boundary layer, the thickness of the first mesh layer is $2.3\times 10^{-5}$m, which yields $y^+\approx$ 1. The error in $\eta_j$ and $\sigma$, obtained from RANS, are less than 5\% and 6\% respectively, compared to those obtained analytically through the Thwaites method. The acoustic mesh is a combination of triangular and quadrilateral mesh, with $\sim$54000 elements. The maximum element size is $4.8\times 10^{-4}$m and the minimum element size is $2.1\times 10^{-5}$m. In the generation of the acoustic mesh, we follow the general rules that the acoustic boundary layer mesh is generated within 4-5 cells. The acoustic boundary layer thickness, $\delta_A = \sqrt{2\mu/\omega \rho}$, is frequency dependent and we ensure there are at least 6-7 cells that within the smallest wavelength. 

\subsection{Experimental setup and procedure\label{sec:experiments}}
\begin{figure}[htbp!]
	\centering
	\includegraphics[width=\columnwidth]{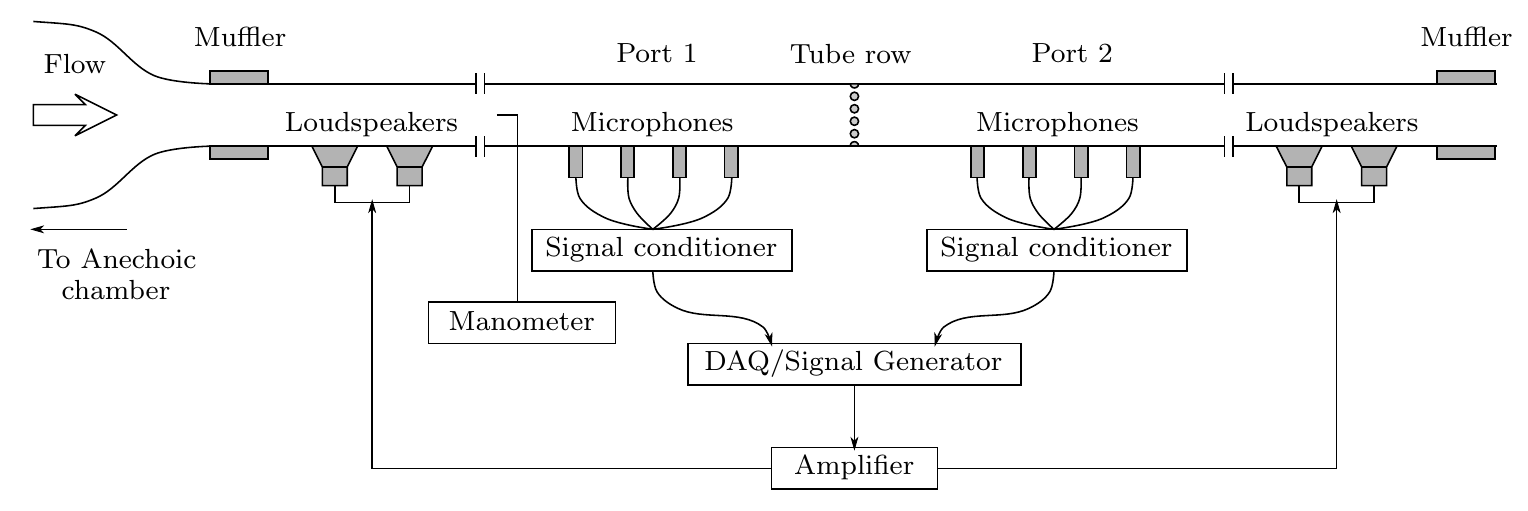}
	\caption{Schematic for the experimental test facility at KTH\label{fig:test_rig}}
\end{figure}

Experiments were conducted at Marcus Wallenberg Laboratory for Sound and Vibration Research (MWL), KTH; the schematic of the flow test rig is as shown in Figure~\ref{fig:test_rig}. It consisted of a long aluminium duct of rectangular cross-section (120mm$\times$25mm) with a wall thickness of 15mm. This was connected to an anechoic chamber at the upstream end and a muffler at the downstream end, in order to reduce acoustic reflections from the duct ends. Acoustic excitation was provided by two pairs of loudspeakers, placed near the  two ends of the duct and far from the tube row sample. This was done to ensure that only plane waves are initially incident on the tube row.

Eight flush mounted microphones (B\&K 4938), four on either side of the sample were used for measuring the up and downstream pressure fluctuations. These microphones were all calibrated relative to each other using a calibrator by subjecting all the microphones to the same sound field. The microphones were connected to the DAQ system (NI C-Series Digitizer Module Cards) through signal conditioners (Nexus conditioning amplifier, Type 2690-A-0S4). The DAQ system also acted as the signal generator for the loudspeakers, who were driven by amplifiers (Lab Gruppen 1600) and with a variable amplification that was adjusted to generate a preferred amplitude in the duct at all frequencies. The upstream flow velocity profile was measured using a static-pitot probe and a SWEMA3000 pressure transducer.

A two-port multi-microphone method \cite{Peerlings2015, Jing1999} was used to obtain the pressure data. Assuming a time dependence of $e^{-i\omega t}$, the measured complex pressure ($p'(x)$) at any position $x$ in the duct can be written in terms of the forward ($p^+$) and backward ($p^-$) travelling waves as
\begin{equation}
p'(x) = p^+ \exp\left(i k^+ x\right) + p^- \exp\left(-i k^- x\right)
\label{eq:pres_data}
\end{equation}
\noindent where $k^+$ and $k^-$ are the forward and backward travelling wavenumbers. The wavenumbers are corrected for the visco-thermal losses as well as the convection effects as \cite{Pierce1994},
\begin{equation}
k^\pm = \frac{k_0 + \left(1+i\right)\beta_0}{1\pm M}
\label{eq:waveno}
\end{equation}
\noindent where
\begin{equation}
\beta_0 = \frac{1}{2\sqrt{2}}\frac{L_p}{S_p}\sqrt{\frac{\omega \nu}{c^2}}\left(1 + \frac{\gamma-1}{\sqrt{Pr}}\right),
\label{eq:beta0}
\end{equation}
\noindent $\nu$ is the kinematic viscosity of air, $L_p$ is the perimeter of the duct cross-section, $S_p$ is the duct cross-sectional area and $Pr$ is the Prandtl number. There are two unknown pressure amplitudes in Eq.~(\ref{eq:pres_data}) namely $p^\pm$, and hence measured pressure data at at least two positions are required to estimate these unknowns. In the experiments, however, pressure data is obtained from four microphones on either side of the sample. This will lead to an overdetermined system as shown in Eq.~(\ref{eq:pres_amp}), which is solved using Moore-Penrose pseudo inverse method \cite{Stewart1990} to obtain the unknown pressure amplitudes $p_{1,2}^\pm$.
\begin{equation}
\left[\begin{matrix}
\exp\left(i k^+ x_{1,2}^1\right) & \exp\left(-i k^- x_{1,2}^1\right)\\
\exp\left(i k^+ x_{1,2}^2\right) & \exp\left(-i k^- x_{1,2}^2\right)\\
\exp\left(i k^+ x_{1,2}^3\right) & \exp\left(-i k^- x_{1,2}^3\right)\\
\exp\left(i k^+ x_{1,2}^4\right) & \exp\left(-i k^- x_{1,2}^4\right)
\end{matrix}\right]\left[\begin{matrix}
p_{1,2}^+\\p_{1,2}^- \end{matrix}\right] = 
\left[\begin{matrix}
p\left(x_{1,2}^1\right)\\
p\left(x_{1,2}^2\right)\\
p\left(x_{1,2}^3\right)\\
p\left(x_{1,2}^4\right)
\end{matrix}\right]
\label{eq:pres_amp}
\end{equation}
\noindent The superscript of $x$ denotes the microphone number, and the subscripts 1 and 2 indicate the upstream and downstream sides, respectively.\\

Once the unknown pressure amplitudes are determined, the scattering coefficients can be calculated from two independent pressure field measurements \cite{Abom1991} as shown in Eq.~(\ref{eq:qs_scat_pres_meas}), where the two pressure fields are generated by independent upstream excitation (denoted by superscript A) and downstream excitation (denoted by superscript B).
\begin{equation}
\left[\begin{matrix}
p_2^{+A} & p_2^{+B} \\ p_1^{-A} & p_1^{-B}
\end{matrix}\right]=\left[\begin{matrix}
T_{1\rightarrow2} & R_2 \\ R_1 & T_{2\rightarrow1}
\end{matrix}\right]\left[\begin{matrix}
p_1^{+A} & p_1^{+B} \\ p_2^{-A} & p_2^{-B}
\end{matrix}\right]
\label{eq:qs_scat_pres_meas}
\end{equation}

This multi-microphone method relies on the acoustic field being one-dimensional. In our experiments, the cross-section is 120mm corresponding to a cut-off frequency of $\sim$1400Hz, beyond which the multi-microphone method breaks down as we encounter transverse modes. Therefore, we restrict our measurements to 1200Hz and the scattering matrix of the tube row was obtained using stepped sine excitations with a frequency step of 50Hz.

\section{Results and Discussion\label{sec:results}}

\begin{figure}[H]
	\centering
	\includegraphics[width=0.875\columnwidth]{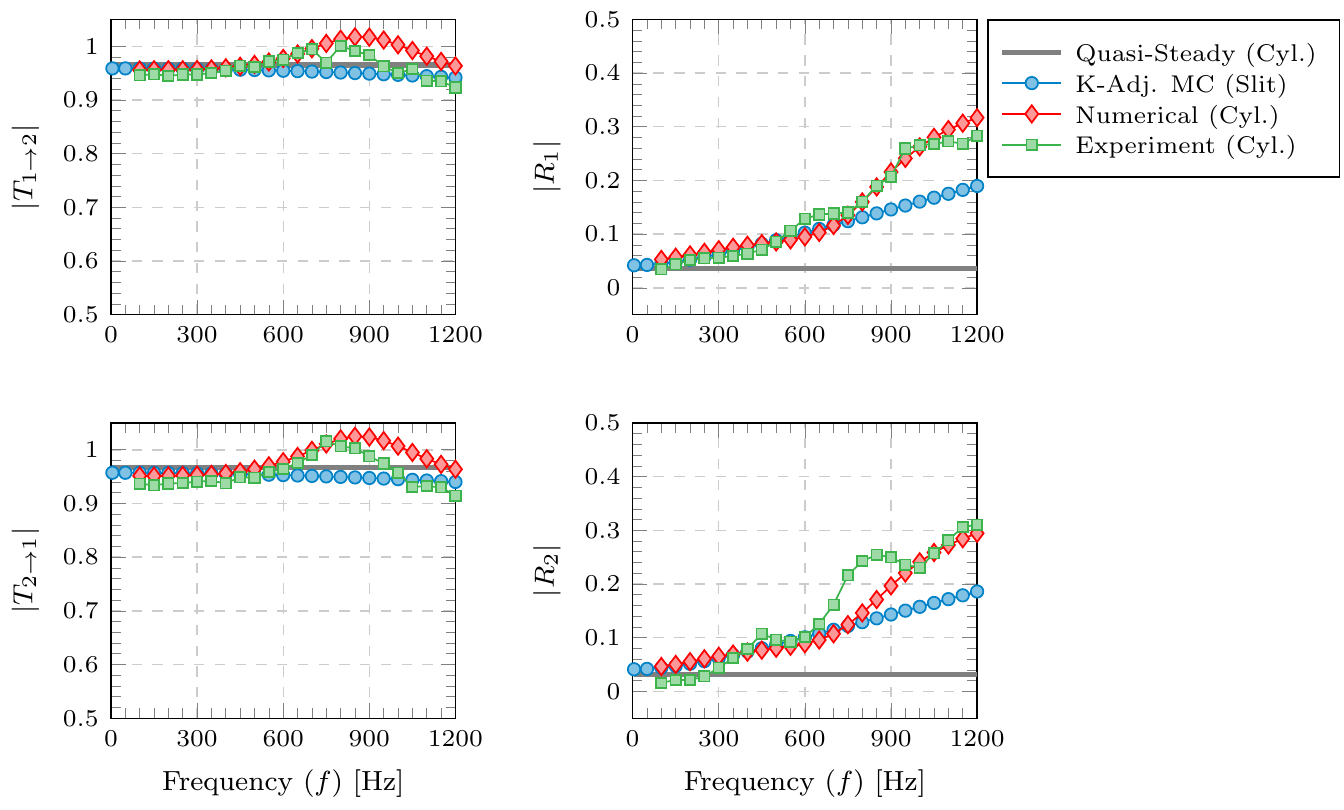}
	\caption{Magnitude of transmission and reflection coefficients for $u_1$ = 6.1m/s.}
	\label{fig:scat_U6_abs}
\end{figure}

\begin{figure}[htbp!]
	\centering
	\includegraphics[width=0.875\columnwidth]{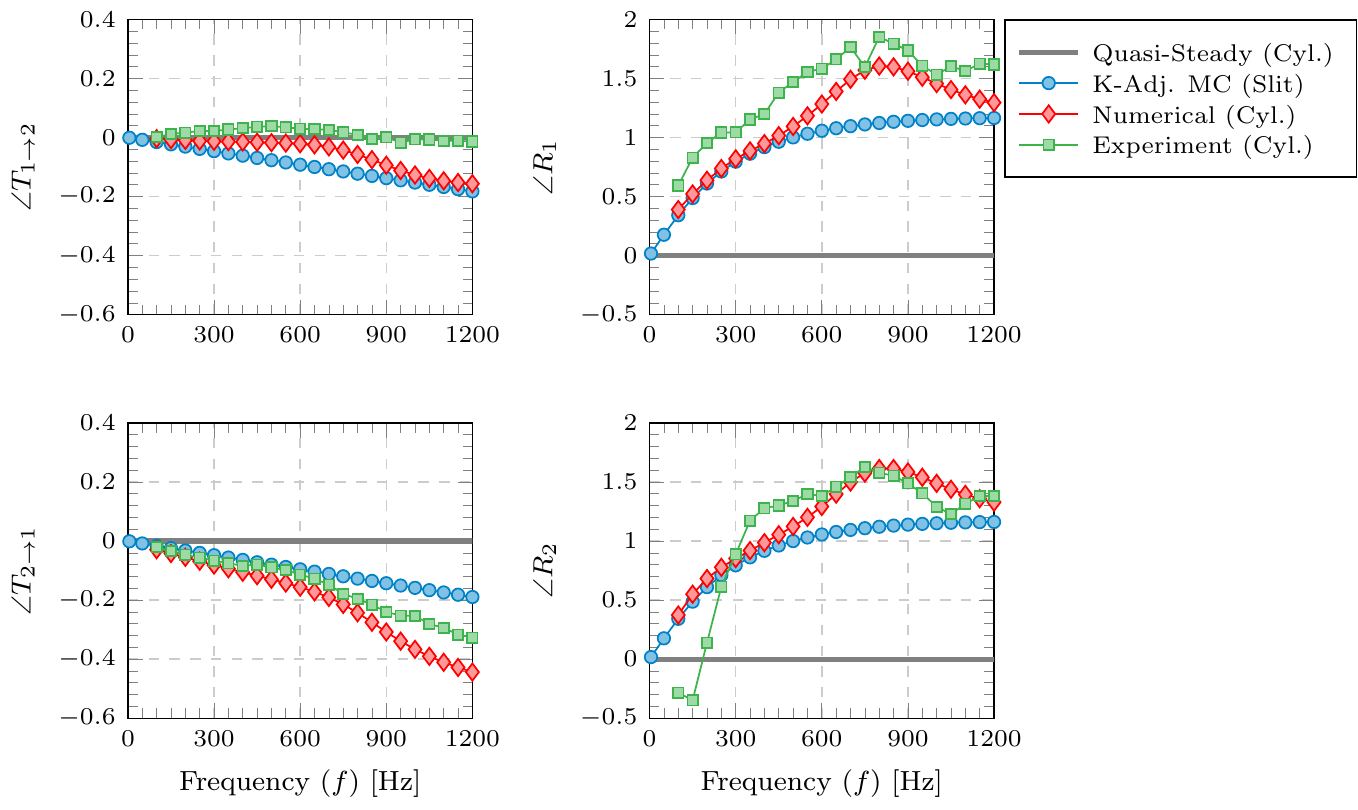}
	\caption{Phase of transmission and reflection coefficients for $u_1$ = 6.1m/s.}
	\label{fig:scat_U6_phs}
\end{figure}

\begin{figure}[H]
	\centering
	\includegraphics[width=0.875\columnwidth]{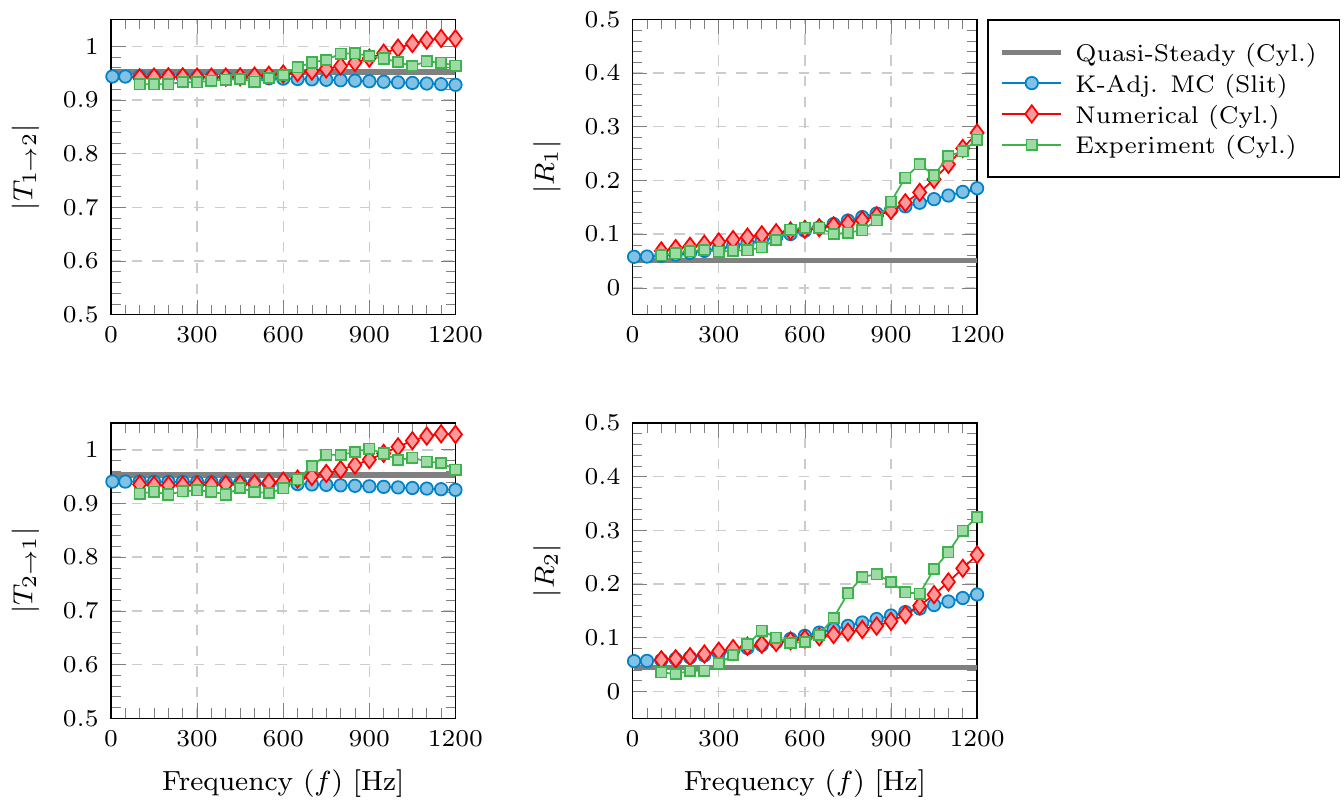}
	\caption{Magnitude of transmission and reflection coefficients for $u_1$ = 8.38m/s.}
	\label{fig:scat_U8_abs}
\end{figure}

\begin{figure}[H]
	\centering
	\includegraphics[width=0.875\columnwidth]{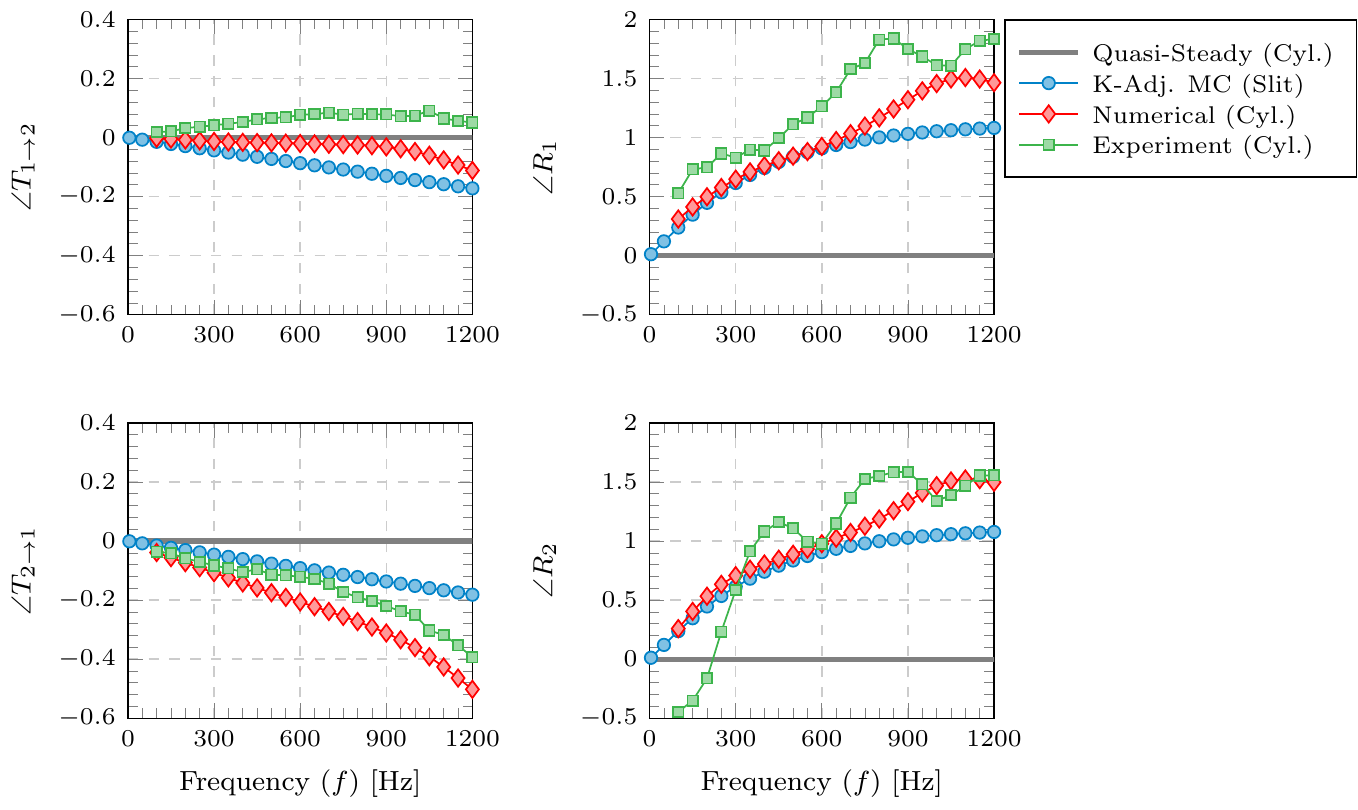}
	\caption{Phase of transmission and reflection coefficients for $u_1$ = 8.38m/s.}
	\label{fig:scat_U8_phs}
\end{figure}

\begin{figure}[H]
	\centering
	\includegraphics[width=0.875\columnwidth]{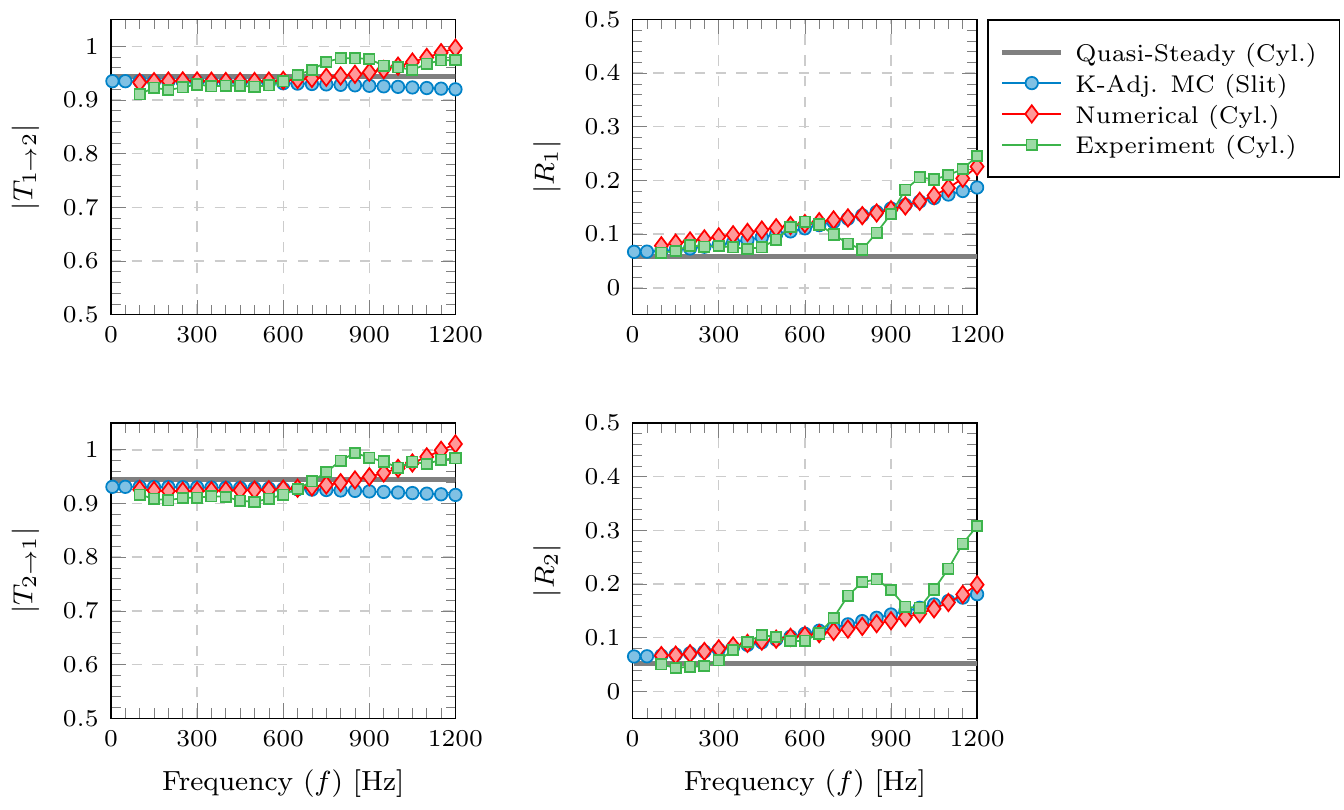}
	\caption{Magnitude of transmission and reflection coefficients for $u_1$ = 9.82m/s.}
	\label{fig:scat_U9_abs}
\end{figure}

\begin{figure}[H]
	\centering
	\includegraphics[width=0.875\columnwidth]{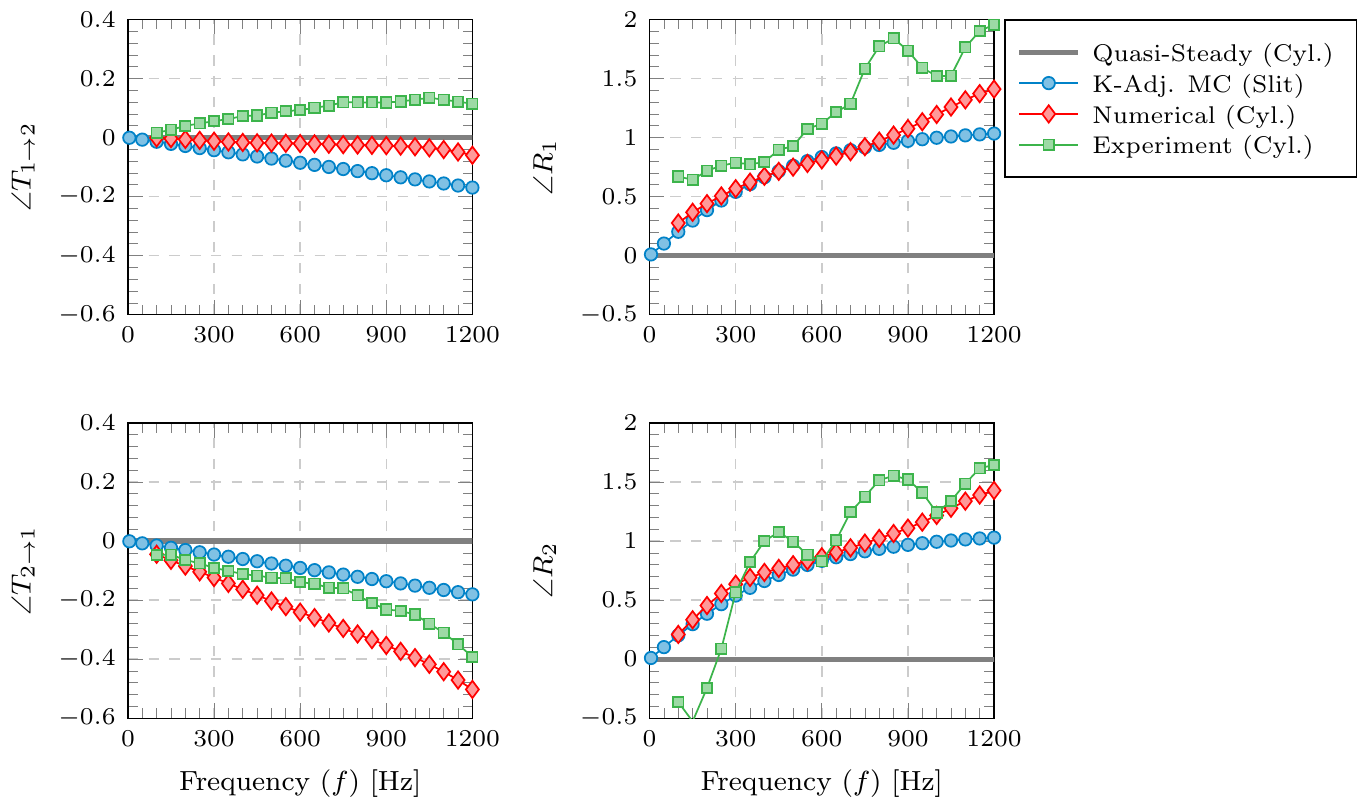}
	\caption{Phase of transmission and reflection coefficients for $u_1$ = 9.82m/s.}
	\label{fig:scat_U9_phs}
\end{figure}

To validate the proposed $K$-adjusted MC model (KMC), scattering matrices were obtained for three inlet velocities: 6.1, 8.38 and 9.82m/s for a tube row of tube diameter $d=11.8$mm with an open-area ratio $\eta=0.31$ and for the frequency range of 100-1200Hz. Results from the QS model, the KMC model, the LNSE predictions and the experimental measurements are shown in Figures~\ref{fig:scat_U6_abs}~-~\ref{fig:scat_U9_phs}. The magnitude of the scattering coefficients for the three different flow speeds are shown in Figures~\ref{fig:scat_U6_abs},~\ref{fig:scat_U8_abs},~\ref{fig:scat_U9_abs} and the phases are shown in Figures~\ref{fig:scat_U6_phs},~\ref{fig:scat_U8_phs},~\ref{fig:scat_U9_phs}. From the magnitude plots we can observe that the QS model, as the name suggests, predicts the quasi-steady behaviour of the tube rows in cross-flow and therefore does not provide information on the frequency dependence of $|T_{1\rightleftarrows2}|$ and $|R_{1,2}|$. For the phase predictions, the QS model assumes that there are no phase changes across the tube row, whereas the KMC model, the LNSE simulations and the experimental measurements indicate phase changes. Nevertheless, the QS model predictions agree well with the results from the KMC model, the LNSE simulations and the experiments in the zero frequency limit.

As the frequency increases, e.g. between 0Hz and 600Hz, the KMC model predictions for the magnitudes and the phases are seen to agree well with the LNSE predictions and the experiments, making it an improvement over the existing QS model. There is, however, a slight mismatch in the predictions for $\angle T_{1\rightarrow2}$; while KMC predicts a definite while small phase change, LNSE and experiments show that the phase change for low frequencies are almost negligible.


\begin{figure}[H]
	\centering
	\includegraphics[width=0.875\columnwidth]{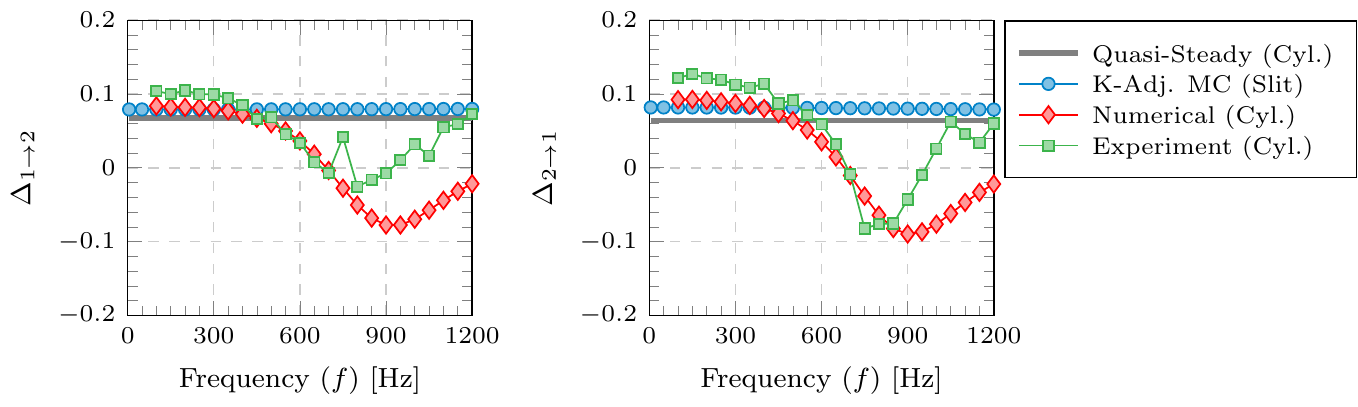}
	\caption{Upstream and downstream absorption coefficients for $u_1$ = 6.1m/s.}
	\label{fig:scat_U6_abscoeff}
\end{figure}

For higher frequencies ($\sim$ 600Hz and above), the LNSE simulations and the experiments predict ``peaks'' in $|T_{1\rightleftarrows2}|$ and $|R_{1,2}|$ that are not predicted by the KMC model. Moreover, $|R_{1,2}|$ display certain ``waviness'' in the experimental results that are captured neither by the KMC model nor by the LNSE simulations. A plausible explanation for this behaviour could be the vortex formation and their convection downstream of the tube row \cite{Peters1993, Moers2017}. However, the waviness peaks do not shift in frequency as we vary the incoming velocity i.e., there seems to be no scaling with $St$ which is typical for hydrodynamic dominated phenomena. Also, contributions from other acoustic and vibrational sources inherent to the test rig cannot be ruled and this needs to be further investigated. Similarly, there are discrepancies between the phases predicted by the KMC model and those obtained from the  experiments and LNSE. As the inlet velocity increases, the frequency at which the transmission coefficients peak shifts to higher values, indicating a $St$ dependence for this behaviour. This may be due to the tube row generating rather than damping acoustic energy, and can be confirmed by plotting the upstream and downstream absorption coefficients (see Figure~\ref{fig:scat_U6_abscoeff}), where negative values for absorption coefficients indicate acoustic energy generation. Since there is both acoustic energy production and a $St$ dependence for the peaks in the scattering coefficients, one is inclined to attribute this behaviour to a more complex acoustic-vortex interaction mechanism in this $St$ range such as that studied in \cite{Yang2016}. Such an acoustic energy contribution is not accounted for in the KMC model and is beyond the scope of the current work.



\section{Conclusion\label{sec:conclusion}}
In the present paper, we have developed an analytical model based on slit plates that can be used for estimating the low frequency scattering coefficients of tube rows in cross-flow. This was achieved by matching the loss coefficients across the tube row and the slit plate. The model was validated against numerical predictions using linearised Navier Stokes equations as well as through experiments, and the predictions for both magnitude and phase for transmission and reflection coefficients agree well with the simulations and experiments. Our model is an improvement over the main previous model for cylindrical tube rows as it accounts for the frequency dependence and hence phase information of the generated acoustic waves. Also, if loss coefficients of other geometries with similar flow features are known, our model can be used for predicting the scattering behaviour. However, the proposed model is limited as it does not predict acoustic sources at high frequencies, as observed in the comparison with experiments and simulations. For certain frequencies, transmission and reflection coefficients from the simulations and experiments exhibited peaks that were not predicted by the current model. For those frequencies, the absorption coefficients turned negative indicating the presence of an acoustic source. Moreover, the peaks shifted to higher frequency values as the inlet velocity increased, denoting a Strouhal number dependence for this behaviour, possibly a more intricate acoustic-vortex interaction phenomenon. 

\section*{Acknowledgements}
We gratefully acknowledge the financial support from the European Research Council (ERC) Consolidator Grant AFIRMATIVE (2018–2023, Grant Number 772080). We also acknowledge the technical inputs and discussions undertaken with Dr.~Ignacio Duran, Reaction Engines Limited, U.~K., as well as the financial assistance provided by Reaction Engines Ltd.~in carrying out experiments at KTH. We are thankful to Mr.~Shail Shah and Ms.~Charitha Vaddamani, KTH for their assistance with conducting experiments.

\bibliography{surendran_paper_biblio}

\end{document}